\documentclass{ws-procs9x6}

\begin{document}

\title{D\O\ QCD studies}

\author{M. VOUTILAINEN For the D\O\ collaboration\\
University of Nebraska in Lincoln, \\
Helsinki Institute of Physics}

\address{Fermilab --- D\O\ M.S. 357, \\
Batavia, IL 60510-0500, USA\\ 
E-mail: mavoutil@fnal.gov}

\maketitle

\abstracts{A number of recent measurements from D\O\ that can be used to constrain parton distributions and tune QCD Monte Carlo models are presented. The selection includes W charge asymmetry, Z+jet event properties, dijet azimuthal decorrelations and the inclusive jet cross section.}

\section{Introduction}

The production of particle jets in hadron collisions is described by the theory of Quantum Chromodynamics (QCD). Production rates of jets can in some cases be predicted by perturbative QCD (pQCD), but frequently Monte Carlo models are needed. The QCD processes often act as a background for many other processes, and knowledge of the parameters of QCD is one of the leading limitations of beyond Standard Model searches. A number of studies that can constrain global parton distribution fits in pQCD and tune Monte Carlo models are discussed in the following sections.

\section{W charge asymmetry}

A measurement\cite{wcharge} of the $W^{\pm}$ rapidity ($y_{W}$) distributions in $p\bar{p}$ collisions provides useful information about the parton distribution functions (PDF) of the $u$ and $d$ quarks in the proton. In this measurement, the resonant production of the $W$ boson constrains the momentum transfer $Q^2$ to $\approx M_W^2$, where $M_W$ is the mass of the $W$ boson. Hence the region in phase space in $x$ that this measurement can probe depends on the range of the rapidity of the $W$ boson. At the center-of-mass energy $\sqrt{s}=1.96$~TeV, this measurement probes the region in $x$ between 0.005 and 0.3.

\begin{figure}[ht]
\centerline{
  \includegraphics[width=0.6\textwidth]{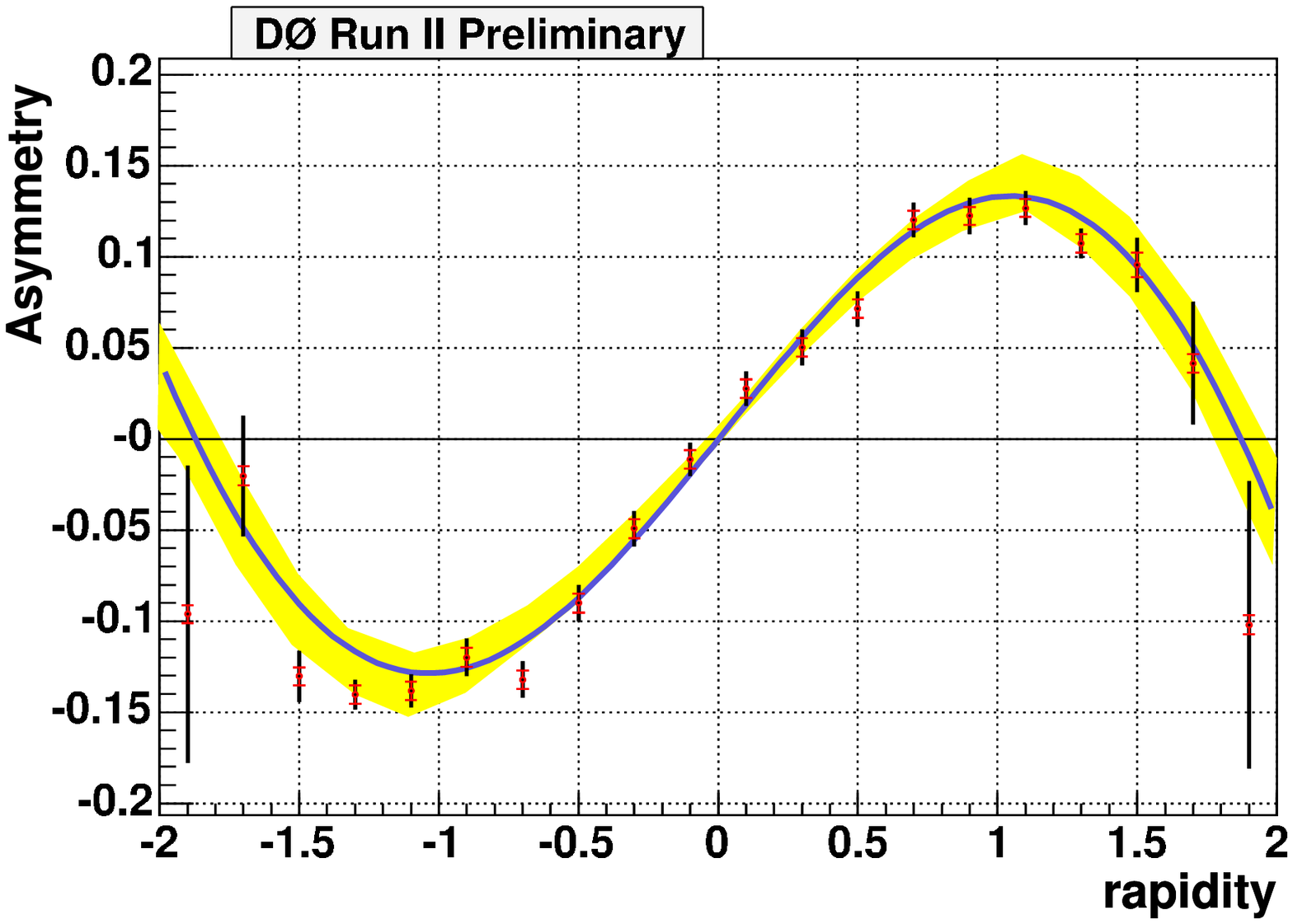}
  \includegraphics[width=0.4\textwidth]{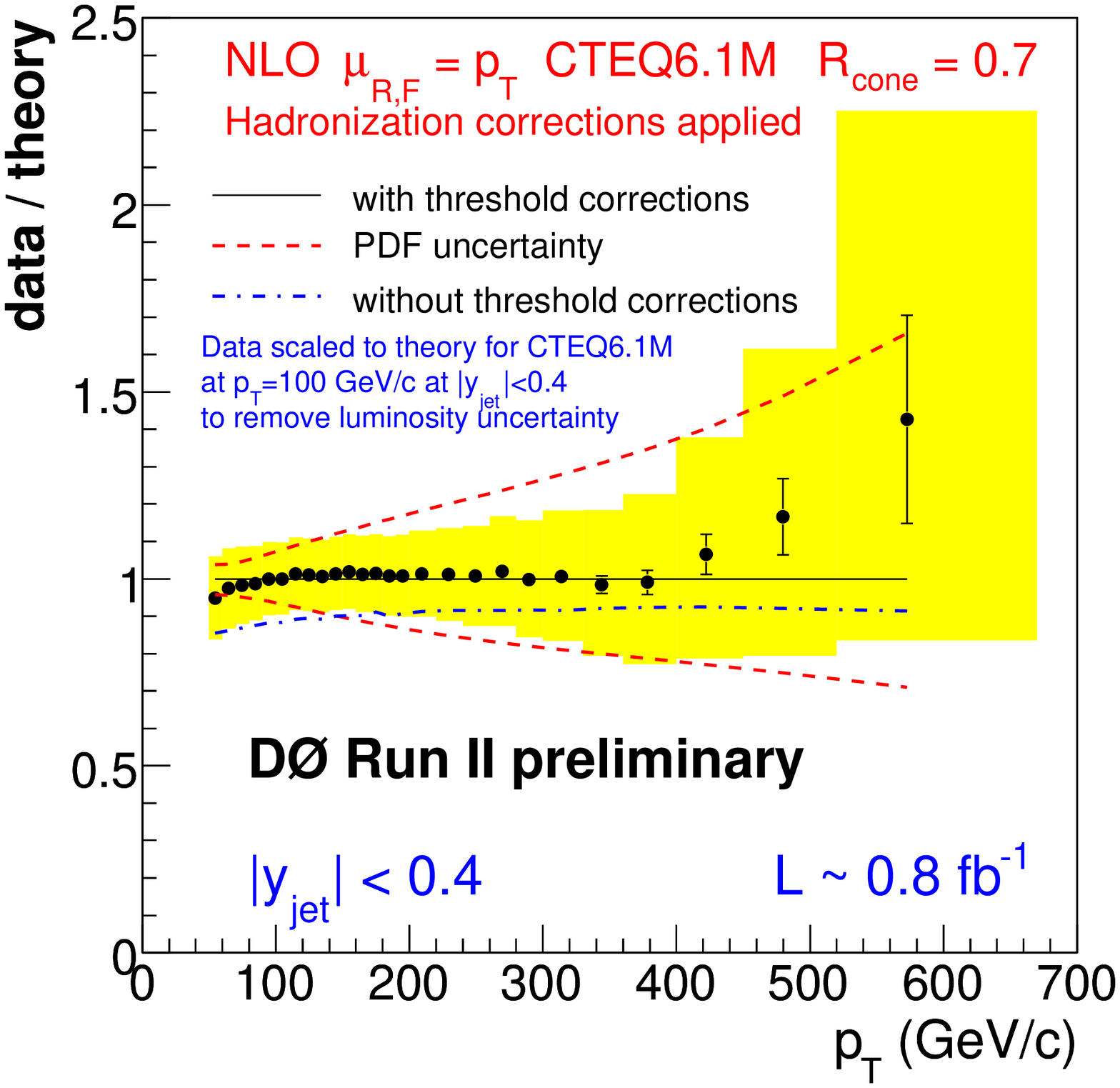}
}
\caption{Muon charge asymmetry distribution with the outer error bars statistical and inner systematic. The error band is the CTEQ6.1M PDF uncertainty and the central curve is the asymmetry using the MRST02 PDF (left). Inclusive jet production cross section compared to theory with the systemic errors as a band and statistical errors as bars (right).\label{fig:wcharge}\label{fig:incljet}}
\end{figure}

The $W$ bosons at the Tevatron are primarily produced by quark-antiquark annihilation. The main production processes are $u+\bar{d}\rightarrow W^+$ and $d+\bar{u}\rightarrow W^-$. Because $u$ valence quarks carry on average more of the momentum of the proton than $d$ valence quarks, the $W^+$ boson is boosted along the proton beam direction and the $W^-$ boson along the antiproton beam direction, giving rise to the $W$ production charge asymmetry
\begin{equation}
A(y) = \frac{d\sigma(W^+)/dy-d\sigma(W^-)/dy}{d\sigma(W^+)/dy+d\sigma(W^-)/dy} \approx \frac{d}{u}.
\end{equation}
It is difficult to measure the $W^\pm$ rapidity due to the fact that the longitudinal momentum of the neutrino from the $W$ decay cannot be measured. Instead, we access the same information by measuring the charge asymmetry of the $W$ boson decay products. In this analysis we use the muon decay channel. The muon asymmetry, shown in Fig.~\ref{fig:wcharge}, is a convolution of the $W$ production charge asymmetry and the asymmetry from the ($V-A$) decay that is well understood. 
The results are compared to NLO prediction with MRST02\cite{mrst} and CTEQ6.1M\cite{cteq6} PDFs with associated errors.
At $\approx 230$~pb$^{-1}$ of data collected, the measurement is not yet systematics limited, which bodes well for this analysis as more data collected by the D\O\ detector is analyzed.

\vspace{-0.2in}

\section{Inclusive jet cross section}

The inclusive jet cross section\cite{incljet} in $p\bar{p}$ collisions at large $p_T$  is directly sensitive to the strong coupling constant  ($\alpha_s$) and the PDFs of the proton. Furthermore, potential deviations from the  pQCD prediction at high $p_T$, not explained by PDFs, may indicate new physics beyond the Standard Model.

The results shown in Fig.~\ref{fig:incljet}(right) are reaching a precision that allows one to reduce the gluon density uncertainty at high momentum fraction. This measurement is covered in more detail in the Hadronic Final States part of these Proceedings.

\section{Dijet azimuthal decorrelations}

Multi-parton radiation is one of the more complex aspects of perturbative Quantum Chromodynamics. The proper description of radiative processes is crucial for a wide range of precision measurements as well as for searches for new physical phenomena where the influence of QCD radiation is unavoidable.

A way to study radiative processes is to examine their impact on angular distributions\cite{azimuth}, shown in Fig.~\ref{fig:azimuth}(left). Dijet production in hadron-hadron collisions, in the absence of radiative effects, results in two jets with equal transverse momenta with respect to the beam axis ($p_{T}$ and correlated azimuthal angles $\Delta\phi_{\rm{dijet}}=|\phi_{\rm{jet1}}-\phi_{\rm{jet2}}|=\pi$. Additional soft radiation causes small azimuthal decorrelations, whereas $\Delta\phi_{\rm{dijet}}$ significantly lower than $\pi$ is evidence of additional hard radiation with high $p_{T}$. Exclusive three-jet production populates $2\pi/3<\Delta\phi_{\rm{dijet}}<\pi$ while smaller values of $\Delta\phi_{\rm{dijet}}$ require additional radiation such as a fourth jet in an event. The results are well-described in pQCD at next-to-leading order in the $\alpha_{s}$, except at large azimuthal differences where soft effects are significant.

\section{Z+jets event properties}

\begin{figure}[ht]
\centerline{
  \includegraphics[width=0.372\textwidth]{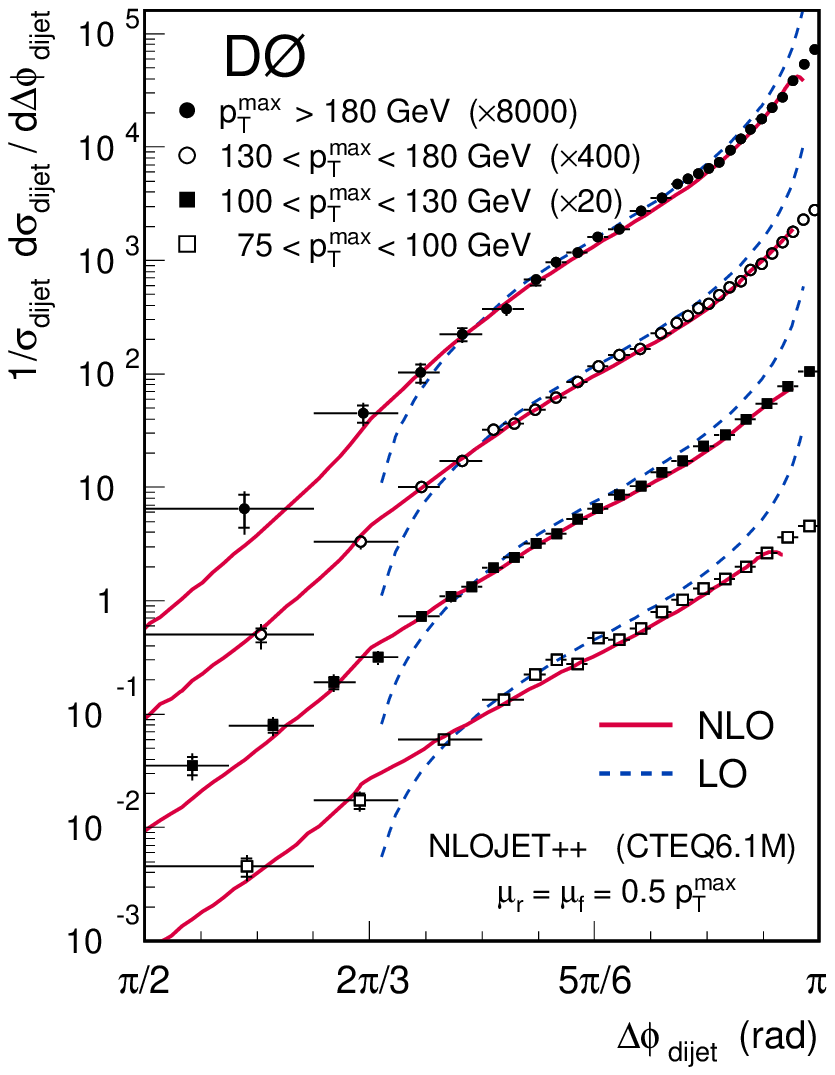}
  \hspace{0.25in}
  \includegraphics[width=0.605\textwidth]{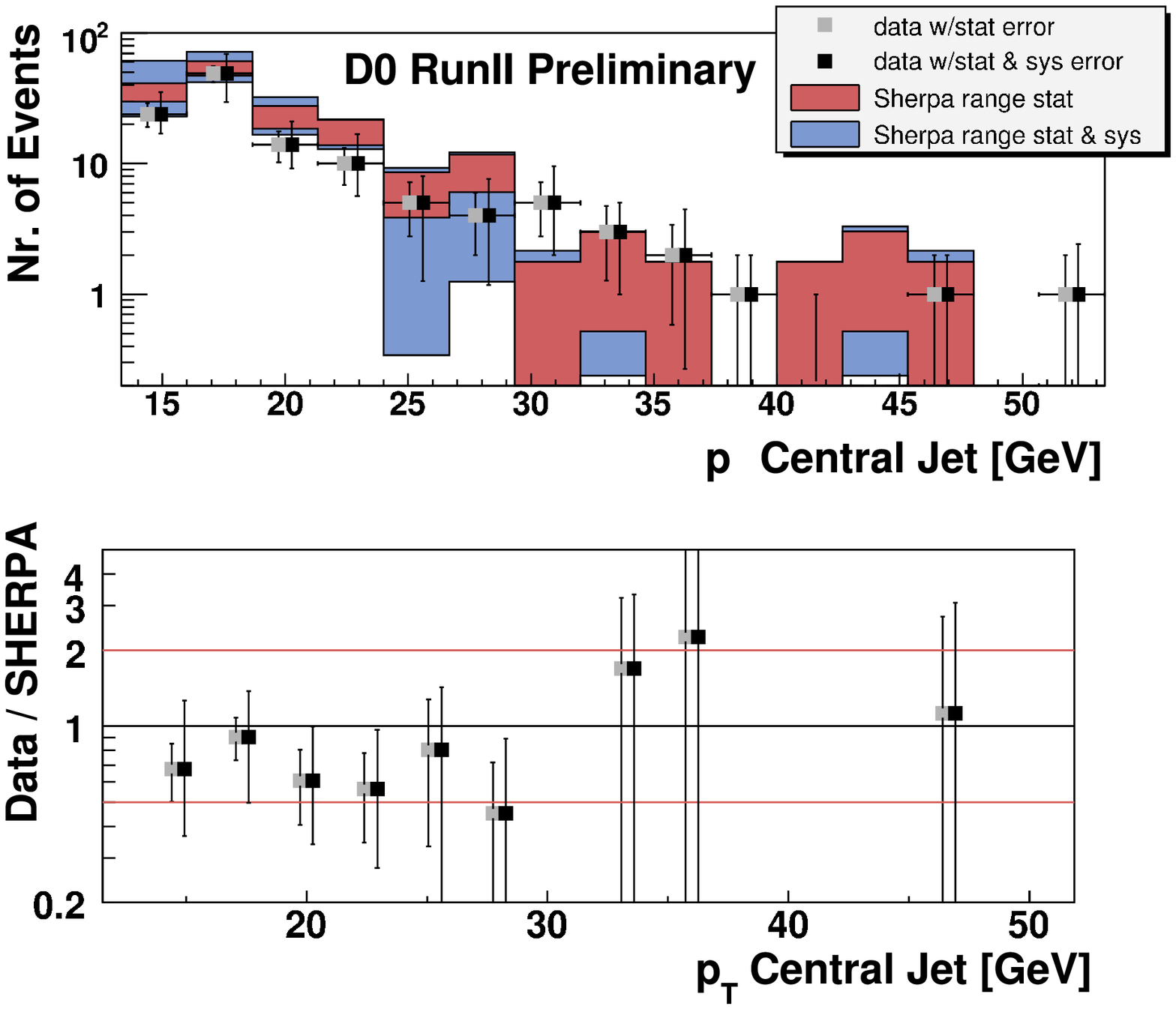}
}
\caption{The $\Delta\phi_{\rm{dijet}}$ distribution in four regions of $p_{T}^{\rm{max}}$ (left). The $p_{T}$ of the third jet lying in between two hardest jets in $\eta$ in $Z$+jet jet events (right).\label{fig:azimuth}\label{fig:zjets}}
\end{figure}

The study\cite{zjet} of the associated production of a vector boson with jets represents an important test of QCD at hadron colliders. In addition, $W/Z$+jet production constitutes an important background in the search for many new physics processes, e.g. in the Higgs boson search in the associated $WH$ and $ZH$ production. The most accurate approach to describe multi-particle final state would include all particles in a full matrix element computation including all real and virtual diagrams. To make calculations tractable, they can be simplified to use a parton shower approach on a $2\rightarrow 2$ core process, or matrix element calculations without virtual corrections. During the last few years several partially overlapping approaches for combining the two methods have been proposed, one of them being the CKKW algorithm. The SHERPA event generator offers an implementation of the CKKW\cite{catani,krauss} algorithm.
In this study, the prediction from the SHERPA Monte Carlo, combining parton shower and matrix element calculations, has been found to give an accurate description of jet multiplicities. In addition, the $p_T$ spectra of the $Z$ boson and of the jets, as shown {\it e.g.} by Fig.~\ref{fig:zjets}(right), as well as angular correlations between jets are reasonably well described.

\vspace{-0.2in}
\vspace{-0.2in}

\section{Conclusion}

Results on $W$ charge asymmetry, dijet azimuthal decorrelations, $Z$+jet event properties and inclusive jet cross section were presented. General agreement between perturbative QCD predictions, Monte Carlo models and data is found. These measurements provide useful feed-back for Monte Carlo models and with improving precision can be used to constrain PDFs.

\end{document}